\documentclass[12pt]{iopart}
\usepackage{iopams}
\expandafter\let\csname equation*\endcsname\relax
\expandafter\let\csname endequation*\endcsname\relax
\usepackage{amsmath}
\usepackage{mathrsfs}
\usepackage{float}
\usepackage{graphicx,color}
\usepackage[english]{babel}
\usepackage{cite}
\usepackage[bookmarks=true,colorlinks,citecolor=blue]{hyperref}

\begin{document}

\newcommand\bra[1]{\mathinner{\langle{\textstyle#1}\rvert}}
\newcommand\ket[1]{\mathinner{\lvert{\textstyle#1}\rangle}}
\newcommand\braket[1]{\mathinner{\langle{\textstyle#1}\rangle}}
\newcommand\hata{\hat{a}}
\newcommand\hatb{\hat{b}}
\newcommand\hatc{\hat{c}}
\newcommand\hatC{\hat{C}}
\newcommand\hatH{\hat{H}}
\newcommand\hatL{\hat{L}}
\newcommand\hatN{\hat{N}}
\newcommand\hatU{\hat{U}}
\newcommand\hatV{\hat{V}}
\newcommand\hatW{\hat{W}}
\newcommand\calC{\mathcal{C}}
\newcommand\calD{\mathcal{D}}
\newcommand\calH{\mathcal{H}}
\newcommand\calP{\mathcal{P}}
\newcommand\calS{\mathcal{S}}
\newcommand\calZ{\mathcal{Z}}
\newcommand\calL{\mathcal{L}}
\newcommand\msc[1]{\textcolor{red}{\textbf{MSC}: #1}}
\definecolor{gray}{gray}{0.6}

\title{Self-Purification and Entanglement Revival in Lambda Matter}



\author{Dongni Chen}
\address{Department of Physics, Korea University, Seoul 02841, South Korea}

\author{Stefano Chesi}
\address{Beijing Computational Science Research Center, Beijing 100193, People's Republic of China}
\address{Department of Physics, Beijing Normal University, Beijing 100875, People's Republic of China}

\author{Mahn-Soo Choi}
\address{Department of Physics, Korea University, Seoul 02841, South Korea}
\ead{choims@korea.ac.kr}

\begin{abstract}
In this study, we explore the dynamics of entanglement in an ensemble of qutrits with a lambda-type level structure interacting with single-mode bosons. Our investigation focuses on zero-energy states within the subspace of totally symmetric wave functions. Remarkably, we observe a universal two-stage dynamics of entanglement with intriguing revival behavior. The revival of entanglement is a consequence of the self-purification process, where the quantum state relaxes and converges universally to a special dark state within the system.
\end{abstract}

%
%
%
%
%

\section{Introduction}

Quantum entanglement plays a pivotal role in quantum information processing, specifically in quantum communication~\cite{Bennett1993_PRL70-1895,Ekert1991_PRL67-661}, quantum simulations~\cite{Seth1996_Science273-5278}, and quantum computing \cite{Bennett2000_Nature404-247}. However, environmental noise poses a significant challenge to preserving entanglement and coherence. For the majority of systems, decoherence results in an irreversible loss of entanglement, becoming a major hindrance in advancing quantum information technologies~\cite{Zurek2003_RMP75-715}.
Despite this common behavior, certain exceptional systems exhibit a fascinating phenomenon known as \emph{entanglement revival}, where entanglement increases during their dissipative evolution. Moreover, a related scenario called `entanglement sudden birth' has been explored, wherein entanglement arises from a separable state after a specific time interval.
Entanglement revival and sudden birth have been extensively studied in a variety of discrete and continuous variable systems~\cite{Zyczkowski2002_PRA65-012101,Aolita2015_RPP78-042001,RTanas2004_JPB6-S90, Benatti2006_JPA39-2689, Cunha2007_NJP9-237, AbdelAty2008_JPB41-235503, Ficek2006_PRA74-024304, Ficek2008_PRA77-054301, Paz2008_PRL100-220401,Drumond2009_JPA42-285308,Das2009_JPB42-205502,Orszag2010_AOP-229,Nunavat2023_MPLA-2350056,Lakhfif2022_PLA-128247,Mirko2018_PRA98-042325}.
This non-monotonic behavior relies on a delicate interplay between entanglement generation, induced by internal system interactions, and dissipation from the external environment \cite{Zyczkowski2002_PRA65-012101, Aolita2015_RPP78-042001, Benatti2006_JPA39-2689}.
Generally speaking, the entanglement revival shows a sensitive dependence on the initial state~\cite{Ficek2006_PRA74-024304, Ficek2008_PRA77-054301, Paz2008_PRL100-220401,Orszag2010_AOP-229,Nunavat2023_MPLA-2350056}.
In some systems, it was suggested that revivals might be induced by the transfer of entanglement between different subsystems~\cite{Yonac2007_JPB40-S45, Lopez2008_PRL101-080503, Bai2009_PRA80-044301}.

In this work, we present a comprehensive investigation of a robust mechanism for entanglement revival, centered around the utilization of special metastable states. These unique states act as entanglement reservoirs during the later stages of time evolution and can be deliberately engineered by leveraging specific (exact or approximate) symmetries inherent in the underlying quantum system. Our focus lies on an ensemble of qutrits exhibiting a lambda-type level structure interacting with a single bosonic mode, a model pertinent to various platforms such as
cavity quantum electrodynamics (QED)~\cite{HarocheRMP2001,EsslingerRMP2001}, trapped ions~\cite{MonroeRMP2021},  circuit QED~\cite{BlaisRMP2021},
and atomic ensembles~\cite{Laurat2007_PRL99-180504}.
The dissipative evolution in our study follows a universal two-stage dynamics. In the initial stage, we observe the expected rapid decay of entanglement at short times. Subsequently, in the second stage leading to entanglement revival, the system undergoes a slower evolution caused by relaxation, ultimately converging to a special dark state.
This self-purification process plays a crucial role in restoring entanglement. Importantly, the entire process is closely connected to an antisymmetry (rather than symmetry) associated with the parity operator of the system.
Furthermore, we provide a comprehensive description of the quantum states involved, enabling a clear identification of the underlying physical processes at play. Armed with this in-depth understanding, we anticipate that the essential features of the revival dynamics can be realized in a broader class of systems.

This paper is structured as follows: In Section \ref{Sec:Model}, we introduce the system, detailing its fundamental physical processes and associated properties. Section \ref{Sec:Entanglement revival} explores the observation of entanglement revival behavior, illustrating the dissipative evolution characterized by universal two-stage dynamics. Section \ref{Sec: semiclassical description} provides a semiclassical perspective on the system, while Sections \ref{Section:Boson loss} and \ref{Section:Qutrit decay} elaborate on the detailed decay processes based on this semiclassical description. Finally, Section \ref{Sec:Conclusion} summarizes the findings and presents an outlook for future research.

\section{Model} 
\label{Sec:Model} 
\begin{figure}
\centering
\includegraphics[width=0.7\textwidth]{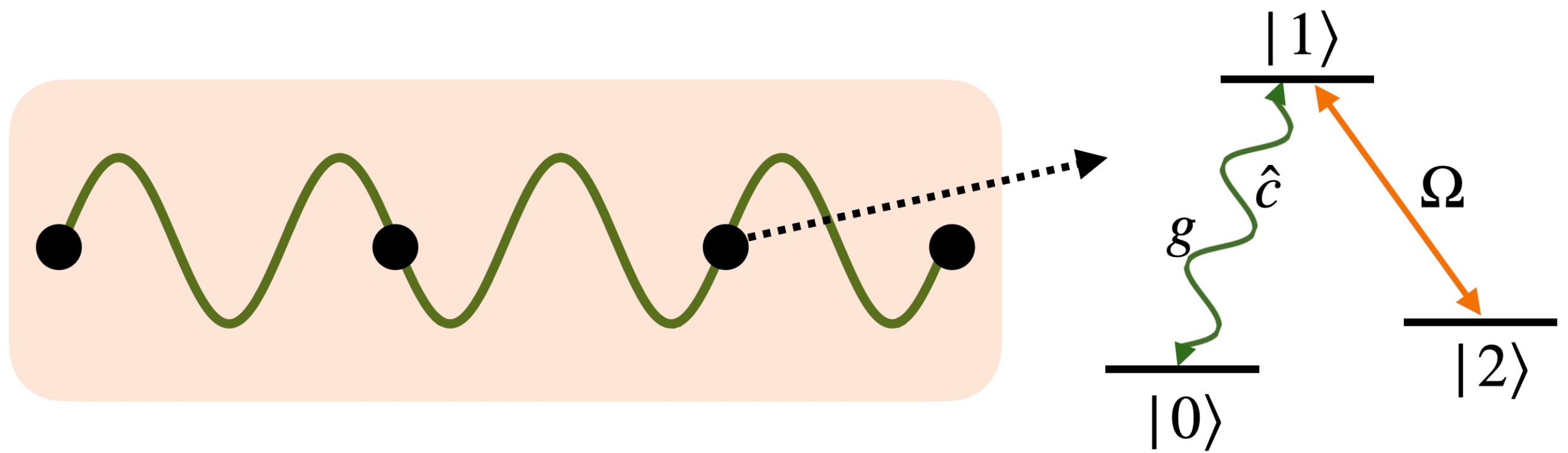}
\caption{Left: Schematics of an ensemble of qutrits (dots) uniformly coupled to a single bosonic mode. Right: The level structure of each qutrit.}
\label{model}
\end{figure}

We consider $n$ identical three-level systems (qutrits, for short) coupled with single-mode bosons. As shown in Fig.~\ref{model}, each qutrit has a $\Lambda$-type structure with two ground-state levels, denoted by $|0\rangle$ and $|2\rangle$, and one excited-state level, $|1\rangle$. The bosonic mode induces the transition $|{0}\rangle\leftrightarrow|1\rangle$ whereas
the transition $|{1}\rangle\leftrightarrow|{2}\rangle$ is driven resonantly by an external classical field. In the interaction picture, the system is governed by the Hamiltonian
\begin{equation}
\hatH = g\sum_{k=1}^n\hatc\,|1\rangle_k\langle0|
+ \Omega\sum_{k=1}^n|1\rangle_k\langle2| + \mathrm{h.c.} ,
\label{Hamiltonian:1}
\end{equation}
where $\hat{c}$ is the bosonic annihilation operator, $|\mu\rangle_k$ ($\mu=0,1,2$) are the quantum states of the $k$th qutrit, $g$ is the qutrit-boson coupling, and $\Omega$ is the Rabi transition amplitude. Note that, in this work, we assume uniform $g$ and $\Omega$ for all qutrits. The operator
\begin{math}
\hat{N} = \hat c^\dag \hat c +
\sum_{k=1}^n(|1\rangle_k\langle 1| + |2\rangle_k\langle 2|)
\end{math} 
is conserved by $\hat H$, and we will usually refer to its eigenvalues (i.e., the total number of excitations) as $p$.

Any realistic system is subject to the influence of quantum noise, thus its dynamics is not unitary any longer. Dissipative dynamics is typically modeled by a quantum master equation, which we take of the following form
\begin{equation}
\label{LindbladME}
\dot{\hat\rho}(t) =
-i[\hatH,\hat\rho]
+ \kappa\mathcal{L}[\hatc]\hat\rho
+ \Gamma_0\calL[\hatL_0]\hat\rho
+ \Gamma_2\calL[\hatL_2]\hat\rho ,
\end{equation}
where $\hat\rho$ is the density operator of the system and $\mathcal{L}[\hat L]\hat\rho$ is the Lindblad superoperator, defined by $\mathcal{L}[\hat L]\hat\rho:=\hat L\hat\rho\hat L^\dagger-(\hat L^\dagger\hat L\hat\rho+\hat \rho\hat L^\dagger\hat L)/2$ for the linear operator $\hatL$ associated with a quantum decoherence process.
Specifically, $\calL[\hatc]$ is responsible for the loss of bosons while $\calL[\hatL_\mu]$, with
\begin{math}
\hatL_\mu := \sum_{k=1}^n{|{\mu}\rangle_k}{\langle{1}|},
\end{math}
describes the collective spontaneous decay of qutrits from the excited-state level $|{1}\rangle$ to the ground-state level $|\mu\rangle$ ($\mu=0,2$); $\kappa$ and $\Gamma_\mu$ are the corresponding rates.
As implied by the specific form of the quantum master equation~\eqref{LindbladME}, we mainly concentrate on \emph{collective} decay of qutrits. However, as we discuss briefly later and demonstrate in detail in the \ref{Appendixb}, our main results are not affected qualitatively by \emph{individual} qutrit decay.

In this work, we investigate an unusual dynamics of entanglement in the presence of quantum decoherence.  The effect of our interest is most pronounced in the zero-energy subspace of $\hat H$ with $p$ excitations and totally symmetric wave functions (with respect to the exchange of qutrits). Therefore we mainly focus on this subspace, denoted as $\calZ_p$.
Interestingly, it was pointed out~\cite{Chen22a} that the $\calZ_p$ subspace features a decoherence-free nature \cite{Facchi2002_PRL89-s080401,Beige2000_PRL85-1762,Lidar1998_PRL81-2594}, and is always degenerate regardless of the parameters, allowing one to geometrically manipulate quantum states within the subspace.
This property is due to three symmetry properties of the system: the conservation of total number of excitations $\hat N$, the exchange symmetry of qutrits,
and the \emph{anti}-symmetry,
\begin{math}
\{\hat\Pi_1,\hatH\} = 0,
\end{math}
of
\begin{math}
\hat\Pi_1 := \exp\left(i\pi\sum_{k=1}^n|1\rangle_k\langle1|\right).
\end{math}

We denote the basis states of $\calZ_p$ by
\begin{math}
\ket{Z_p^i}
\end{math}
for $i=0,1,2,\cdots,[p/2]$
\footnote{For the dimension of $\calZ_p$, see Ref.~\cite{Chen22a}.},
where $[x]$ indicates the integer part of the real number $x$. The first zero-energy eigenstate can always be chosen of the following form:
\begin{equation}
\label{degeneracy:eq:4}
|Z_{p}^0\rangle
= \sum_{k=0}^{p}
\frac{(-1)^{p-k}}{\sqrt{(p-k)!}g^{p-k}
  \Omega^{k}}|\Phi_n^k\rangle_Q|p-k\rangle_c,
\end{equation}
where
\begin{math}
\ket{\Phi_n^k}_Q
= \sum_\calP\calP\ket{2}^{\otimes k}\ket{0}^{\otimes(n-k)}
\end{math}
($\mathcal{P}$ are permutations of qutrits) is a symmetric Dicke state and
$\ket{k}_c$ is a Fock state with $k$ bosons. Equation~(\ref{degeneracy:eq:4}) describes a special zero-energy state which we call \emph{master dark state}. It has several interesting properties useful for quantum-state engineering applications, including the generation of arbitrary symmetric Dicke states~\cite{Chen22a,Shao2010_EPL90-50003}.
We will see below that this master dark state plays a key role in the entanglement-revival behavior as well.

\begin{figure}
\centering
\includegraphics[width=70mm]{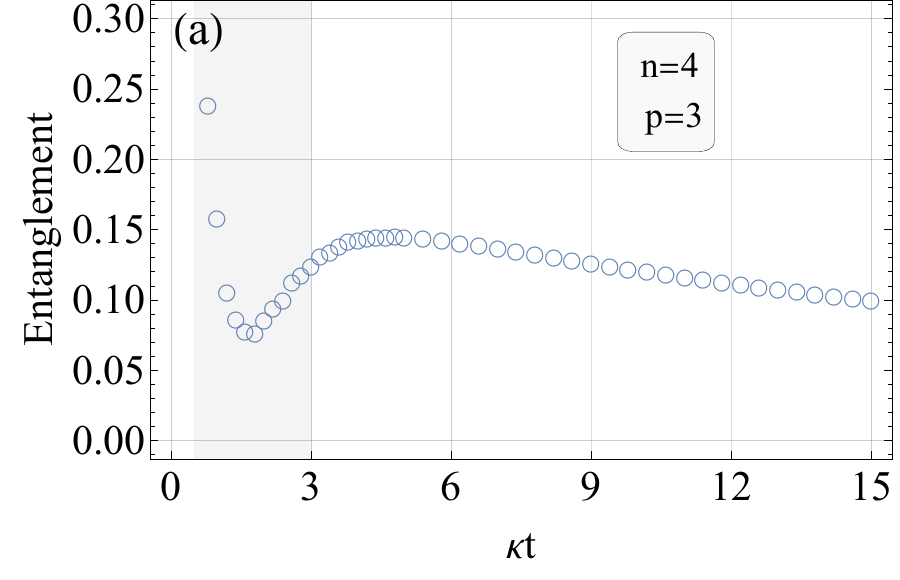}
\hspace{0.2cm}       
\includegraphics[width=70mm]{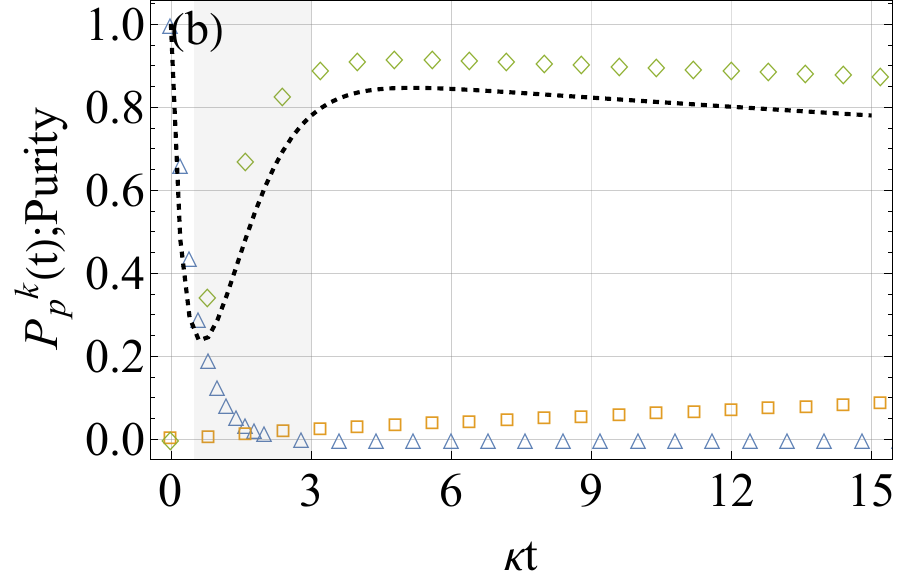}      
\caption{(a): Time dependence of the bipartite entanglement between the bosonic and qutrit subsystems, for $n=4$ and the intial state $|Z_3^1\rangle$ ($p=3$). (b): Probabilities of relevant zero-energy eigenstates,
$\braket{Z_3^1|\hat\rho(t)|Z_3^1}$ (blue triangles), 
$\braket{Z_1^0|\hat\rho(t)|Z_1^0}$ (green diamonds),
$\braket{Z_0^0|\hat\rho(t)|Z_0^0}$ (orange squares), and purity $\Tr\left[\hat\rho^2\right]$ of the density matrix (dashed curve).
We used: $\Omega/g=0.15$, $\kappa/g=0.1$, $\Gamma_{0}/g=0.05$, and $\Gamma_{2}/g=0.0025$.}
\label{Fig2} 
\end{figure}

\section{Entanglement revival}
\label{Sec:Entanglement revival}

Figure~\ref{Fig2} illustrates the main phenomenon in the focus of this work. For four qutrits ($n=4$) with three excitations ($p=3$), which we take as a prototypical example, 
there are two zero-energy states, $\ket{Z_p^0}$ and $\ket{Z_p^1}$, where $\ket{Z_p^1}$ has a relatively large qutrit-boson entanglement.
In this work, we quantify the entanglement by the logarithmic negativity~\cite{Plbnio2007_QInfoCompu7-1,PhysRevLett.95.090503,Jens1999_JModopt46-1}, defined as
\begin{equation}
E_N(\rho_{AB}):=\log_2||\rho_{AB}^{T_B}||,
\end{equation}
where $\|\cdots\|$ denotes the trace norm and $\rho^{T_B}$ indicates the partial transposition of $\rho$ with respect to subsystem $B$.
From the quantum dissipative dynamics of Eq.~\eqref{LindbladME}, we find in Fig.~\ref{Fig2}(a) that the entanglement content of $\ket{Z_p^1}$ decreases fast initially, as usually expected. Surprisingly, however, the entanglement increases again to a certain level as time evolves. This type of two-stage dynamics and entanglement-revival behavior are universally found in a wide range of parameters.

In search for clues to explain the above behavior, we consider the probabilities
\begin{math}
P_p^k(t) := \bra{Z_p^k}\hat\rho(t)\ket{Z_p^k}
\end{math}
of being in a zero-energy state ($p\leq 3$).
We find that, among all zero-energy states, the key role is played by
$\ket{Z_3^1}$, $\ket{Z_1^0}$, and $\ket{Z_0^0}$, whose populations are shown in in Fig.~\ref{Fig2}(b).
In the first stage of the dynamics, the initial state $|Z_3^1\rangle$ decays rapidly (over a time scale of order $1/\kappa$ or $1/\Gamma_\mu$) to the $p=1$ master dark state $|Z_1^0\rangle$.
Once the master dark state $|Z_1^0\rangle$ becomes dominant, the second stage of dynamics kicks in, and $\ket{Z_1^0}$ decays to the trivial state
\begin{math}
\ket{Z_0^0}\equiv\ket{0}_1\otimes\cdots\otimes\ket{0}_n\otimes\ket{p}_c.
\end{math}
The latter process is much slower, since the master dark states are extremely robust to decoherence~\cite{Chen22a}. As inferred from Eq.~\ref{degeneracy:eq:4}, they are not affected by spontaneous decay of qutrits (neither individual or collective) from the excited-state to ground-state levels, and become immune to boson losses in the limit of large $g/\Omega$. 
Between the two stages there is a certain transition period (shaded region in Fig.~\ref{Fig2}) characterized by a low purity, see the dashed curve in Fig.~\ref{Fig2}(b).
Importantly, the probabilities
\begin{math}
P_3^1, P_1^0, P_0^0
\end{math}
do not sum up to unity during this transition period. This means that some other states give a significant contribution to $\hat\rho(t)$. Indeed, as we discuss in detail below, the decay of the initial state $\ket{Z_3^1}$ to $\ket{Z_1^0}$ occurs via intermediate states with finite energies,
accounting for the missing population $1-P_3^1-P_1^0-P_0^0$.
The important point here is that the mixture of those states dramatically suppresses the overall entanglement to a level that is much smaller than the entanglement of each component (pure) state.

We conclude that the entanglement revival can be attributed to an initial suppression of entanglement due to strong mixture, followed by a recovery of entanglement when the system evolves towards the (relatively) pure master dark state
which, like the initial state, is significantly entangled.
This self-purification is possible because the master dark state is stable against the decay processes. As noted already, the effect of boson loss gets suppressed by increasing $g/\Omega$. However, this also leads to a master dark state with smaller entanglement, implying that the visibility of the entanglement revival can be optimized over the coupling ratio in \ref{Appendixa}. 

\section{Semiclassical description}
\label{Sec: semiclassical description}

We now quantitatively analyze the two-stage dissipative dynamics and the
corresponding entanglement-revival behavior, observed numerically above.
To do this, we first simplify our model through a semiclassical approximation,
recalling that we are working in the subspace of totally symmetric wave functions.
As long as the quantum master equation~\eqref{LindbladME} contains only collective decoherence of qutrits and boson loss, the system remains in this subspace.
Then, it is convenient to describe the ensemble of qutrits using bosonic operators $\hata_\mu$ associated with the levels $\ket\mu$.
Expressed in terms of these bosonic operators, $\hat{H}$ reads
\begin{equation}
\label{main::eq:3}
\hatH
= g\hat c^\dagger \hat a_0^\dagger\hat a_1
+ \Omega\hata_2^\dagger\hata_1 + \mathrm{h.c.},
\end{equation}
which preserves the total number of qutrits
\begin{math}
\hata_0^\dagger\hata_0 +
\hata_1^\dagger\hata_1 +
\hata_2^\dagger\hata_2 = n
\end{math}
as well as the total number of excitations
\begin{math}
\hata_1^\dagger\hata_1 +
\hata_2^\dagger\hata_2 +
\hat c^\dagger\hatc = p.
\end{math}
Likewise, Eq.~\eqref{LindbladME} becomes:
\begin{equation}
\label{main::eq:2}
\dot{\hat\rho}(t) =
-i[\hatH,\hat\rho]
+ \kappa\mathcal{L}[\hatc]\hat\rho
+ \Gamma_0\calL[\hata_0^\dag\hata_1]\hat\rho
+ \Gamma_2\calL[\hata_2^\dag\hata_1]\hat\rho.
\end{equation}

So far, everything is exact.
We now make a semiclassical approximation
\begin{math}
g\hat c^\dagger\hat a_0^\dagger\hat a_1
\approx g\hat c^\dagger \hat a_1\sqrt{n},
\end{math}
which is valid in the limit $n\gg p$.
Then, Eq.~\eqref{main::eq:3} gives:
\begin{equation}
\label{main::eq:4}
\hatH \approx \hatH_\mathrm{sc} :=
g\sqrt{n}\,\hatc^\dag\hata_1 +
\Omega\,\hata_2^\dag\hata_1 + \mathrm{h.c.}
\end{equation}
This semiclassical Hamiltonian is quadratic in the bosonic operators, hence can be solved exactly.
By introducing three new bosonic operators
\begin{subequations}
\begin{align}
\hat{C}_{0}  &:= \cos\theta\,\hat{c} - \sin\theta\,\hat{a}_{2}, \\
\hat{C}_{\pm}  &:= \left(\sin\theta\,\hat{c}+\cos\theta\,\hat{a}_{2}
\mp \hat{a}_{1}\right)/\sqrt2,
\end{align}
\end{subequations}
where $\tan\theta:=g\sqrt{n}/\Omega$,
the semiclassical Hamiltonian takes the following form:
\begin{equation}
\label{main::eq:5}
\hatH_\mathrm{sc}
= \epsilon(\hatC_+^\dagger \hatC_+-\hatC_-^\dagger\hatC_-),
\end{equation}
where
$\epsilon:=\sqrt{g^2n+\Omega^2}$. We note from Eq.~(\ref{main::eq:5}) that $\hatC_0$ is a zero-frequency bosonic eigenmode, while $\hatC_\pm$ have opposite mode-frequencies $\pm\epsilon$.
This interesting property is due to the anti-symmetry
\begin{math}
\{\hat\Pi_1,\hatH\} = 0,
\end{math}
with $\hat\Pi_1$ expressed now as
\begin{math}
\hat\Pi_1 = \exp\left(i\pi\hata_1^\dag\hata_1\right).
\end{math}
The semiclassical Hamiltonian immediately yileds the eigenenergies $(k_+-k_-)\epsilon$, with the correspoding eigenstates:
\begin{equation}
\label{main::eq:1}
|E_{k_{0};k_{+}k_{-}}\rangle =
\frac{1}{\sqrt{k_0! k_+! k_-!}}
(\hat{C}_{0}^{\dagger})^{k_{0}}
(\hat{C}_{+}^{\dagger})^{k_{+}}
(\hat{C}_{-}^{\dagger})^{k_{-}}\ket{\;},
\end{equation}
where $\ket{\;}$ is the vacuum state. Of particular importance are the semiclassical zero-energy states with $p$ excitations (corresponding to the
subspace $\calZ_p$), which take the general form 
\begin{math}
\ket{E_{(p-2k);kk}}.
\end{math}
The case $k=0$ gives the master dark states of Eq.~(\ref{degeneracy:eq:4}), i.e.,
\begin{math}
\ket{E_{p;00}} \approx \ket{Z_p^0}.
\end{math}
For $p=3$, applicable to Fig.~\ref{Fig2}, the two semiclassical zero-energy states are
$|E_{3;00}\rangle\approx|Z_3^0\rangle$ and $|E_{1;11}\rangle\approx|Z_3^1\rangle$. 

In the physically relevant regime of $\kappa,\Gamma_\mu\lesssim g,\Omega$, the relative phase between any pair of semiclassical eigenstates with different energies oscillates quickly (the energy difference is of order $ \epsilon\gg\kappa,\Gamma_\mu$).
Therefore, by ignoring the phase coherence between such eigenstates, we consider a semiclassical solution of the following form
\begin{equation}
\label{main::eq:6}
\hat \rho_\mathrm{sc}(t) \approx
\sum_\alpha P_\alpha (t) |E_\alpha\rangle \langle E_\alpha|,
\end{equation}
where $\alpha\equiv(k_0;k_+k_-)$ collectively denotes the semiclassical quantum numbers $k_0$ and $k_\pm$.
Substituting this ansatz into Eq.~\eqref{main::eq:2}, we obtain the (classical) equations
\begin{equation}
\label{semiclassicalME}
\frac{dP_\alpha}{dt} = \sum_{\beta}\gamma_{\alpha\beta}P_\beta(t)
\end{equation}
for the probabilities $P_\alpha(t)$, where 
\begin{math}
\gamma_{\alpha\beta}
\end{math}
is the transition rate between semiclassical states $|\alpha\rangle$ and $|\beta\rangle$.

\begin{figure}
\centering
\includegraphics[width=80mm]{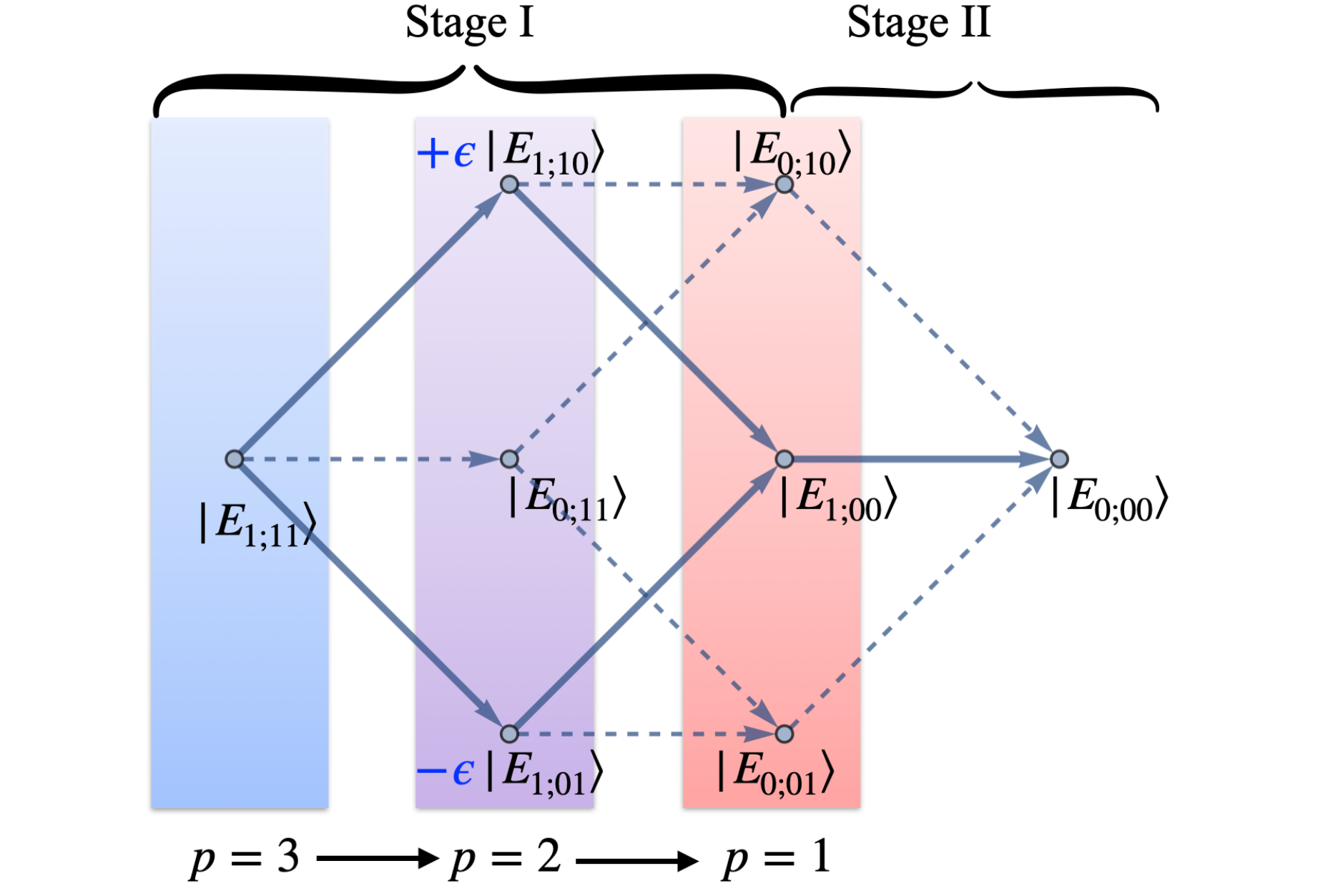}
\caption{Schematic diagram describing boson losses within the semiclassical approximation. Dominant semiclassical processes are represented by solid blue arrows. The complete semiclassical diagram includes both solid and dashed arrows.}
\label{Fig5}
\end{figure}

\section{Boson loss}
\label{Section:Boson loss}

We now apply the semiclassical description of Eqs.~\eqref{main::eq:6} and \eqref{semiclassicalME} to analyze in detail the dynamical process. We first consider the effects of boson loss and, for simplicity, we focus on the particular example with $p=3$. Similar results apply to arbitrary $p$.

In the semiclassical limit ($n\gg p$, hence $\tan\theta\gg 1$), the rates 
\begin{math}
\gamma_{\alpha\beta} := \kappa |\langle E_\alpha|\hat c|E_{\beta}\rangle|^2
\end{math}
of the four processes
\begin{math}
\ket{E_{1;11}} \to \ket{E_{1;10}},
\end{math}
\begin{math}
\ket{E_{1;11}} \to \ket{E_{1;01}},
\end{math}
\begin{math}
\ket{E_{1;10}} \to \ket{E_{1;00}},
\end{math}
and
\begin{math}
\ket{E_{1;01}} \to \ket{E_{1;00}},
\end{math}
are much larger (by a factor of order $g^2n/\Omega^2\gg 1$) then other transition rates.
Therefore, as illustrated in Fig.~\ref{Fig5},
one can identify two dominant (incoherent) decay paths of $\ket{Z_3^1}\approx\ket{E_{1;11}}$:
\begin{align}
\ket{E_{1;11}} \to \ket{E_{1;10}} \to \ket{E_{1;00}}, \\
\ket{E_{1;11}} \to \ket{E_{1;01}} \to \ket{E_{1;00}}.
\end{align}
We also include
\begin{math}
\ket{E_{1;00}} \to \ket{E_{0;00}}
\end{math}
among the major transitions of  Fig.~\ref{Fig5} (solid arrows). The latter process has 
a much smaller rate, but is the only allowed transition once the system has reached $\ket{E_{1;00}}$.

\begin{figure}
\centering
\includegraphics[width=70mm]{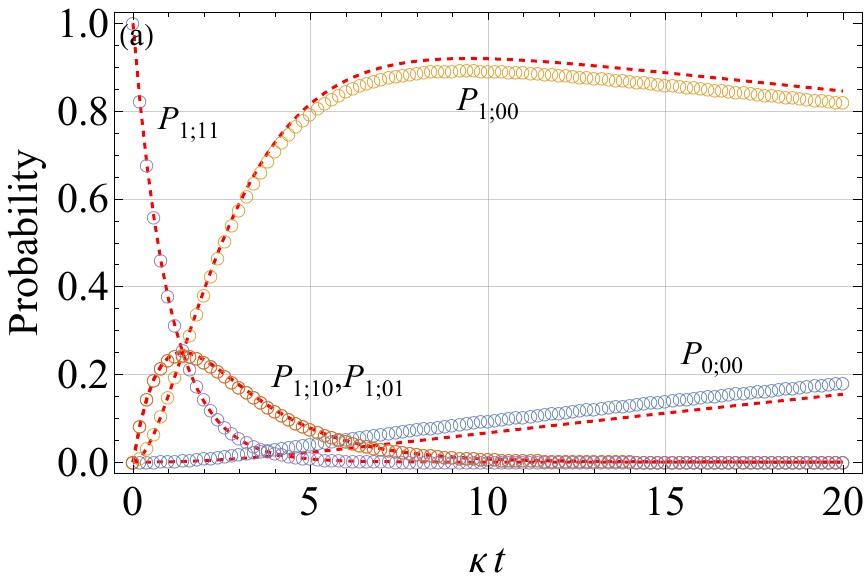}
\hspace{0.2cm}
\includegraphics[width=70mm]{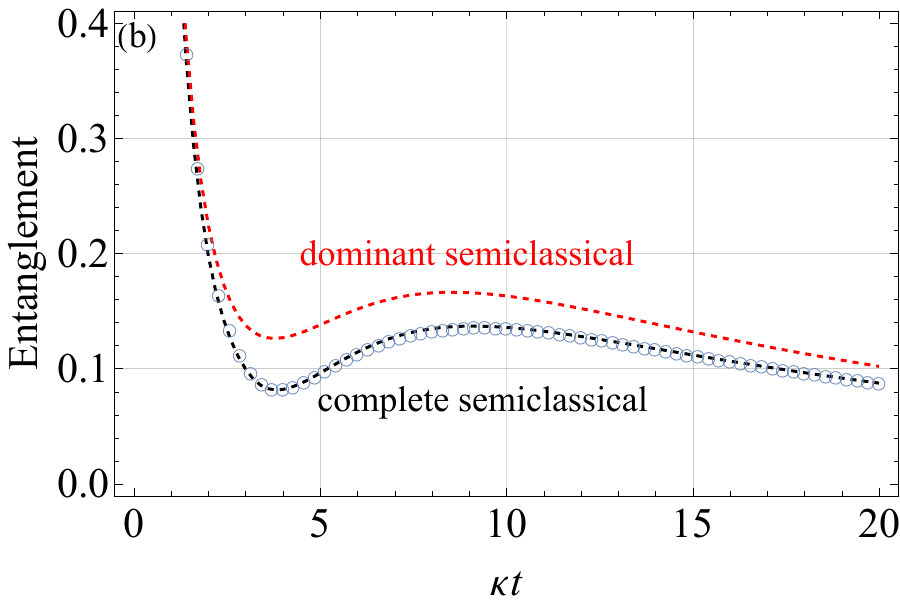}
\caption{Comparison of the exact (empty circles) and semiclassical (dashed curves) dynamics in the presence of boson loss. (a): Probabilities $P_\alpha$ of the five states $|E_\alpha\rangle$ entering the major transition paths of Fig.~\ref{Fig5}. (b): Time dependence of the logarithmic negativity. The dashed curves in panel (a) only consider the dominant semiclassical contribution, while in panel (b) we show both the dominant (upper dashed curve) and the complete (lower dashed curve) semiclassical treatment. The latter includes all states and transitions of Fig.~\ref{Fig5}. In both panels: $n=20$, $p=3$, $\Omega/g=0.1 \sqrt n$, and $\kappa/g=0.1$. } 
\label{Fig6}
\end{figure}

The above remarks allow us to further simplify our semiclassical description, by only including in the ansatz $\hat\rho_\mathrm{sc}(t)$ of Eq.~\eqref{main::eq:6}
the five most relevant states 
\begin{math}
\ket{E_{1;11}},
\end{math}
\begin{math}
\ket{E_{1;10}},
\end{math}
\begin{math}
\ket{E_{1;01}},
\end{math}
\begin{math}
\ket{E_{1;00}},
\end{math}
and
\begin{math}
\ket{E_{0;00}}
\end{math},
which appear in the major transition paths. As Fig.~\ref{Fig6}(a) demonstrates,
the approximate evolution of the five populations $P_\alpha(t)$ agrees very well with the full numerical solution, obtained from Eq.~\eqref{main::eq:2}. 
In Figure~\ref{Fig6}(b) we further compare the entanglement content of the dominant semiclassical state $\hat\rho_\mathrm{sc}(t)$ (red dashed line) to the full solution $\hat\rho(t)$ (blue empty circles).
Here the agreement is worse than for the probabilities of panel (a); apparently, the entanglement content is more sensitive to the detailed  form of $\hat\rho(t)$.
However, the discrepancy can be easily corrected by including all  semiclassical eigenstates $\ket{E_{k_0;k_+k_-}}$ in $\hat\rho_\mathrm{sc}(t)$. In this case, the possible transition paths include those marked by dashed arrows in Fig.~\ref{Fig5}.
The black dashed curve of Fig.~\ref{Fig6}(b) plots the evolution of the entanglement content within the full semiclassical description, and shows an excellent agreement with the full solution. 

We have shown above that, as expected, the semiclassical approximation works very well when $n\gg p$. However, it also gives qualitatively correct predictions of the dynamical behavior for values of $n$ as small as $n=4$ (with $p=3$). Many features can be even described quantitatively, as long as we replace $\hata^\dagger_0\hata_0$ by the average occupation $n_0$ of state $|0\rangle$.

\section{Qutrit decay}
\label{Section:Qutrit decay} 

\begin{figure}
\centering
\includegraphics[width=8cm]{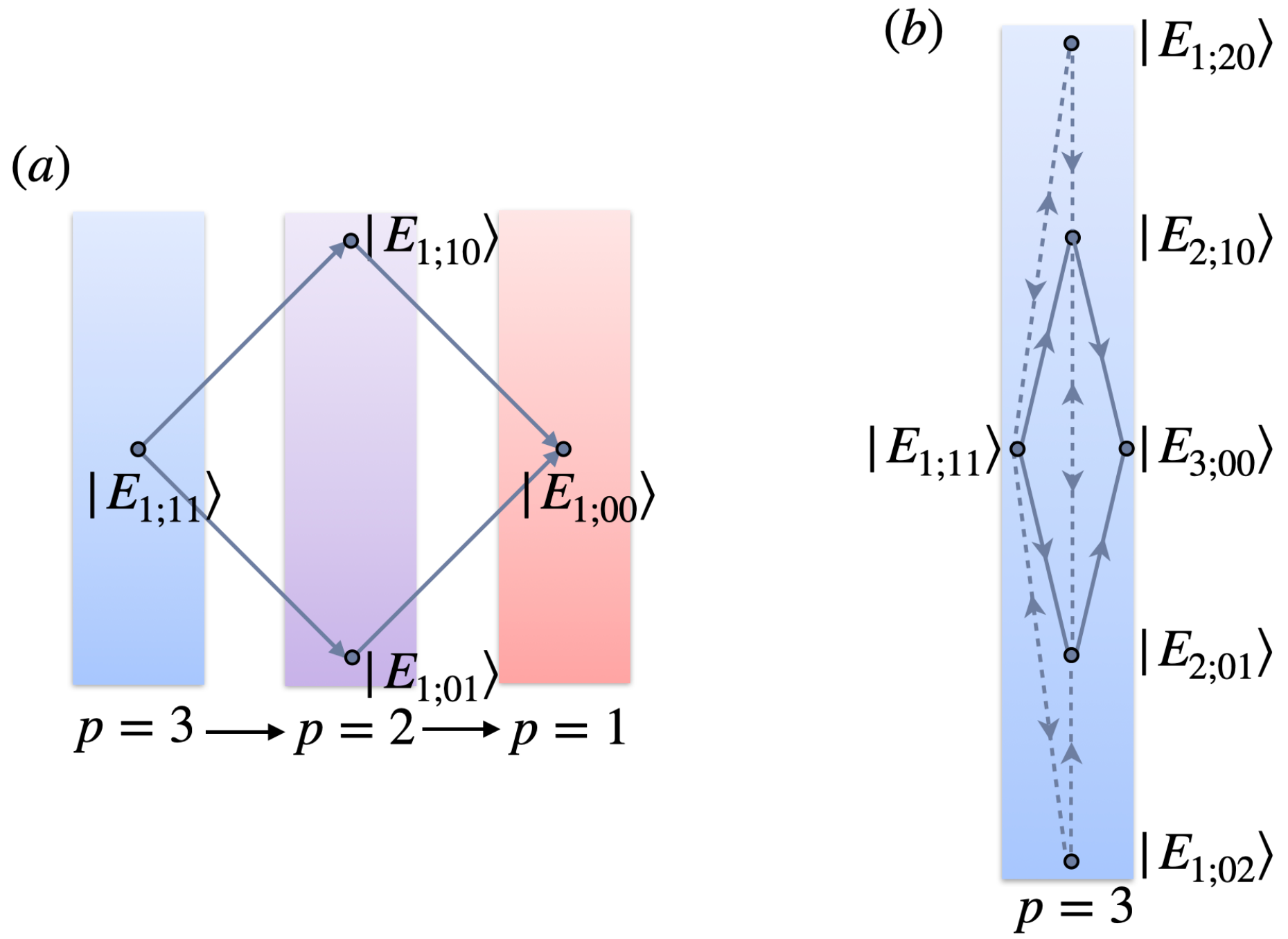}
\caption{Semiclassical diagrams describing the collective decay of qutrits. Panels (a) and (b) refer to $\hata_0^\dagger\hata_1$ and $\hata_2^\dagger\hata_1$ decay, respectively. In panel (b), the rates associated to dashed arrows are suppressed by a factor of order $\Omega^2/(g^2 n) \ll 1$.}
\label{Fig8}
\end{figure}

The semiclassical description is still applicable when collective decay of qutrits is the dominant decoherence mechanism. Assuming  $\kappa=\Gamma_2=0$, we illustrate in Fig.~\ref{Fig8}(a) the transitions induced by the $\hat{a}_0^\dagger \hat{a}_1 \simeq \sqrt{n} \hat{a}_1$ process of Eq.~(\ref{LindbladME}). This semiclassical diagram is similar to the one in Fig.~\ref{Fig5}, except that now mode $\hat{C}_0$ is not affected by dissipation. Therefore, all the slow transitions of Fig.~\ref{Fig5} do not take place (including $|E_{1;00}\rangle \to |E_{0;00}\rangle$). The system will ultimately reach the master dark state $|E_{1;00}\rangle$, instead of the vacuum state. In Fig.~\ref{Fig8}(a), the transition rates are all of order $\Gamma = \sqrt{n} \Gamma_0$, i.e., collectively enhanced by the occupancy of state $|0\rangle$. If, on the other hand, the  $\hat{a}_2^\dagger \hat{a}_1$ process dominates ($\kappa=\Gamma_0=0$) we find the complete semiclassical diagram shown in Fig.~\ref{Fig8}(b). Since the decay from $|1\rangle$ to $|2\rangle$ does not alter the excitation number, the dynamics is confined to the $p=3$ subspace, and terminates in the master dark state $|E_{3;00}\rangle$. 

\begin{figure}
\centering
\includegraphics[width=8cm]{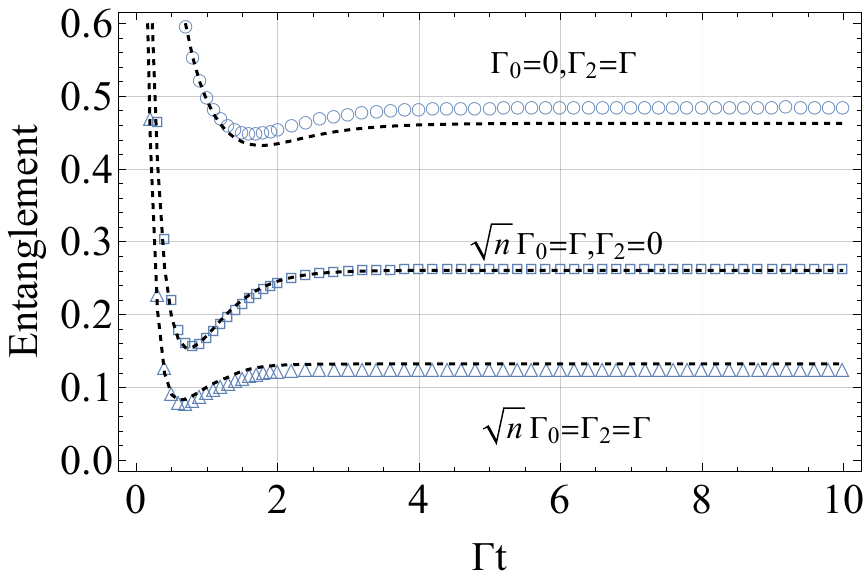}
\caption{Evolution of the logarithmic negativity in the presence of collective qutrit decay. The full solution, obtained from Eq.~\ref{main::eq:2}, is computed with $\Gamma_2=\Gamma$ and $\Gamma_0=0$ (empty circles), $\sqrt{n}\Gamma_0=\Gamma$ and $\Gamma_2=0$ (empty squares), and $\Gamma_2= \sqrt{n}\Gamma_0 = \Gamma$ (empty triangles). The black dashed curves are from the semiclassical approximation. We also used: $n=20$, $p=3$, $\kappa=0$, $\Gamma/g=0.02$, and $\Omega/g=0.1\sqrt n$.}
\label{Fig9}
\end{figure}

As before, both diagrams of Fig.~\ref{Fig8} lead naturally to a two-stage dynamics where, after an initial loss of purity, the system evolves towards a master dark state. As a consequence, revivals are observed in the time dependence of the logarithmic negativity, shown in Fig.~\ref{Fig9}. Since the master dark states are completely immune to the decay of qutrits, at long times the entanglement content saturates to a finite value. 

When both types of qutrit decay are present, the semiclassical diagram is more complex but can be obtained in a similar manner. An entanglement revival is still found with the parameters of Fig.~\ref{Fig9} (see the lowest curve). However, now the system relaxes to a mixture of three master dark states, $|E_{3;00}\rangle$, $|E_{1;00}\rangle$, and $|E_{2;00}\rangle$, leading to to a general reduction of entanglement. As seen, in all three cases of Fig.~\ref{Fig9} the semiclassical approximation (dashed curves) is in close agreement with the exact evolution (empty circles).

Finally, we comment on the effect of individual decay of qutrits. This process is more complex to describe, as it causes transitions between different symmetry sectors, expanding the dynamical process beyond the totally symmetric subspace. Nevertheless, we find that the dynamics is qualitatively similar in \ref{Appendixb}. The system can still reach a distinct master dark state, causing the entanglement to revive. Only when the decay rates from $|1\rangle$ to $|0\rangle$ and from $|1\rangle$ to $|2\rangle$ are comparable, a larger number of states is involved and this might prevent the self-purification process in \ref{Appendixb}. The analysis of individual qubit decay suggests that the revival is robust to other types of local perturbations, e.g., small deviations of $g,\Omega$ from the homogenous limit. The entanglement revival also survives in the presence of detuning in \ref{Appendixc}.

\section{Conclusion}
\label{Sec:Conclusion}

By analyzing the evolution of an ensemble of qutrits interacting with a bosonic mode, we have identified a robust physical mechanism for entanglement revivals.
We show that starting from a highly-entangled state, decoherence leads to a universal two-stage dynamics which can be accurately captured by a semiclassical approximation. The entanglement revival is attributed to the self-purification of the quantum state as it relaxes towards a special dark state.
While the specific model of our study has direct applicability to various experimental setups~\cite{HarocheRMP2001,EsslingerRMP2001,MonroeRMP2021,BlaisRMP2021,Laurat2007_PRL99-180504}, the revival
mechanism boasts broad applicability and bears some analogy to other extensively studied protocols to generate entanglement through dissipation such as reservoir engineering~\cite{Cirac1996,Polzik2011,Wang2013_PRL110-253601,Wang2015_PRA91-013807,Barzanjeh2019}.
However, our primary focus here is on inducing an unconventional form of quantum evolution.

\section*{Acknowledgments}

S.C.\ acknowledges support from the Innovation Program for Quantum Science and Technology (Grant No. 2021ZD0301602), the National Science Association Funds (Grant No. U2230402), and the National Natural Science Foundation of China (Grant Nos. 11974040 and 12150610464).
C.D.\ and M.-S.C.\ has been supported by the National Research Function (NRF) of Korea
(Grant Nos. 2022M3H3A106307411 and 2023R1A2C1005588) and by the Ministry of
Education through the BK21 Four program.

\section*{Data Availability}

Any data that support the findings of this study are included within the article.

\appendix

\section{Optimal coupling ratio}
\label{Appendixa}

Here, we analyze the optimal value of the coupling ratio $\Omega/(g\sqrt{n})$, which is determined by the competing requirements of having a dark state with large entanglement and being weakly affected by boson decay. 

For simplicity and to focus on the essential point, we assume no qutrit decay (the effect of which does not affect the qualitative feature).
In this case, the semiclassical dynamics is described by Fig.~3 of the main text. We also recall that the revival of entanglement is due to a self-purification of the system, as it relaxes to the special master dark state $|E_{1;00}\rangle$. This process takes place on a timescale $\sim\kappa^{-1}$. Later, boson decay causes a slow leakage towards the vacuum state, $|E_{1;00}\rangle \to |E_{0;00}\rangle $. The rate of this process is of order $ \kappa/\tan\theta$ and can be suppressed by making $\tan\theta$ as large as possible. In this limit the probability of the master dark state can approach one (ideal purification), but the entanglement between the qutrits and the cavity becomes vanishingly small. We see from Eq.~(3) of the main text that $|Z_p^0\rangle \simeq |\Phi_n^p\rangle_Q|0\rangle_c$ when $\tan\theta = g\sqrt{n}/\Omega \gg 1$.

\begin{figure}
\centering
\includegraphics[height=55mm]{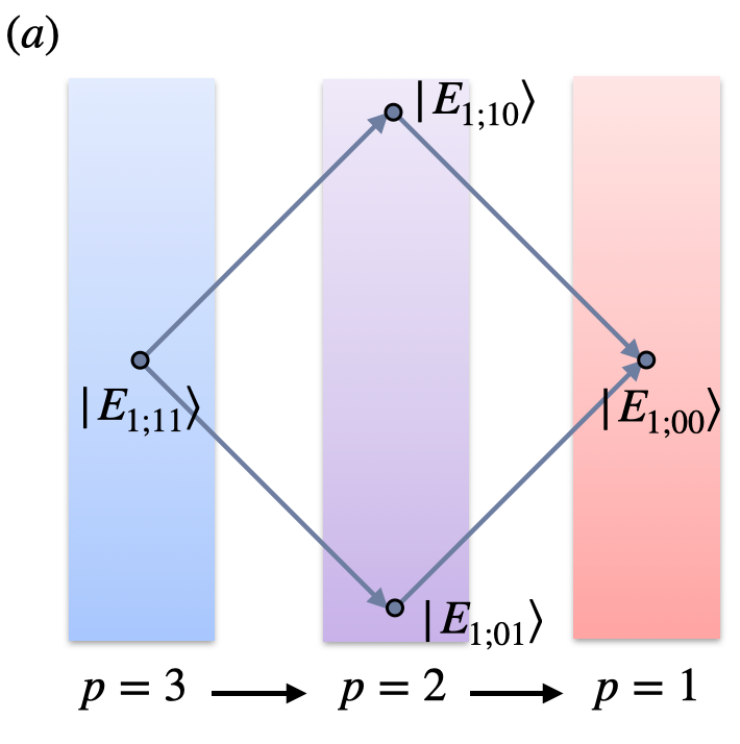} 
\hspace{0.4cm}
\includegraphics[height=55mm]{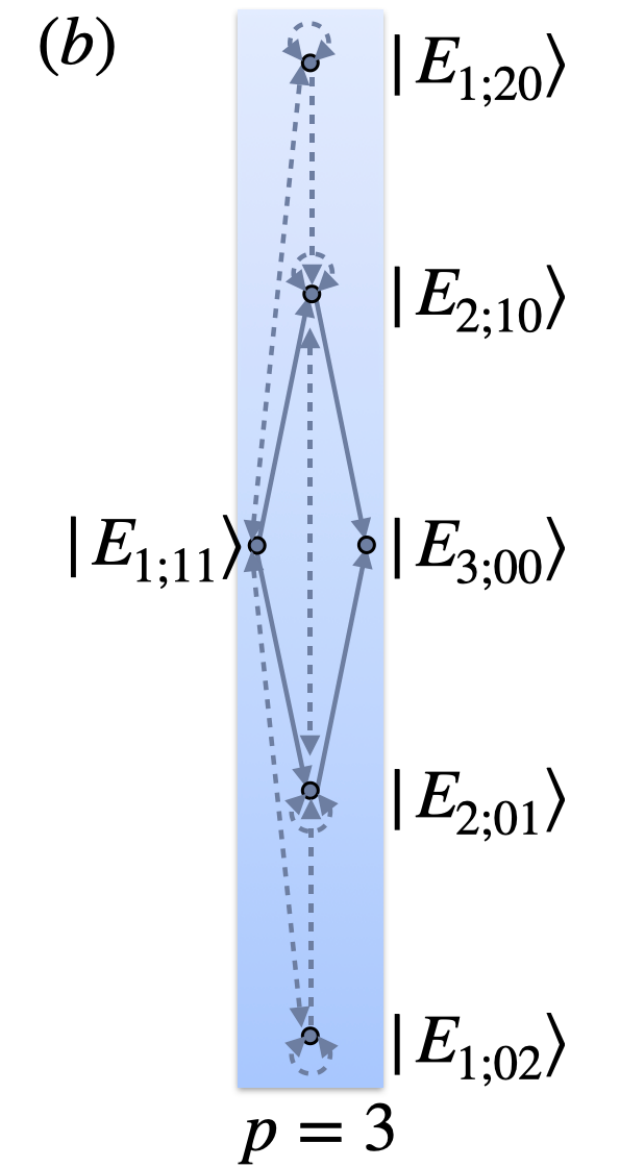}
\caption{(a): Entanglement revivals at different values of the coupling strengths ratio $1/\tan\theta=\Omega/(g\sqrt n)$. The values of $1/\tan\theta$ range from 0.01 to 0.2, in steps of 0.01 (from top to bottom). (b): Revival height as a function of  $1/\tan\theta$. We use $n=20$, $p=3$, $\kappa/g=0.1$ and neglect qutrit decay.}
\label{Supplemental::fig:7}
\end{figure}

Because of this competition,  the entanglement revival will at first become more visible by increasing the ratio of the two coupling strengths $1/\tan\theta=\Omega/(g\sqrt{n})$ but gradually disappears at larger values. This behavior is illustrated in  panel (a) of Fig.~\ref{Supplemental::fig:7}, where we plot the time dependence of the entanglement with various coupling strength ratios. The height of the revival as function of  $1/\tan\theta$ is plotted in Fig.~\ref{Supplemental::fig:7}(b), showing that the optimal ratio is in the range of $0.05-0.1$.

\begin{figure}
\centering
\includegraphics[width=85mm]{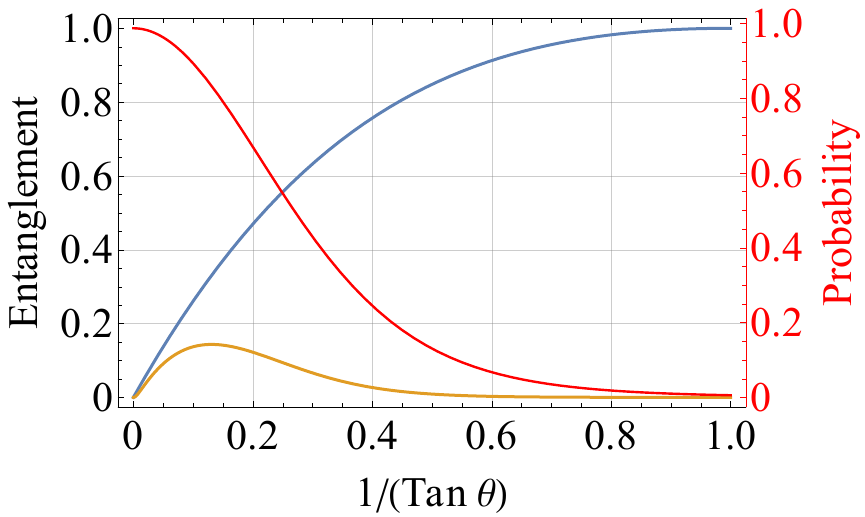}
\caption{Estimation of the optimal coupling ratio. The entanglement of the master dark state  $|E_{1;00}\rangle$ (blue curve) grows with $1/\tan\theta$, while the corresponding probability $P_{1;00}(10)$ (red curve) is a decreasing function of $1/\tan\theta$. The  orange curve is the entanglement of the the approximated density operator $\Tilde{\rho}(10)$. Here $n=20$ and $p=3$.}
\label{Supplemental::fig:8}
\end{figure}

This optimal ratio can be estimated through simple analytical calculations. At long times $t \gg \kappa^{-1}$, the initial state $|E_{1;11}\rangle$ has almost fully decayed to a mixture of $|E_{1;00}\rangle$ and $|E_{0;00}\rangle$. The approximate density matrix, denoted by $\Tilde{\rho} (t)$, can be expressed in the following form:
\begin{equation}
    \Tilde{\rho}(t)=P_{1;00}(t)|E_{1;00}\rangle\langle E_{100}|+(1-P_{1;00}(t))|E_{0;00}\rangle\langle E_{0;00}|,
\end{equation}
 where the probability can be derived through the semiclassical rate equation, Eq.~(11) of the main text, giving
 \begin{equation}
     P_{1;00}(t)=e^{-\kappa t}(1-e^{\frac{\sin\theta}{2(\sin\theta+\cos\theta)}\kappa t})^2.
 \end{equation}
As a reference, we consider the typical timescale $\kappa t=10$ and compute the entanglement of the approximate density operator $\Tilde{\rho}(10)$. The entanglement obtained in this manner is plotted in Fig.~\ref{Supplemental::fig:8}, from where we estimate an optimal coupling ratio $1/\tan\theta\approx 0.13$, which is close to the value from the full simulations of Fig.~\ref{Supplemental::fig:7}. 

\section{Individual decay of qutrits}
\label{Appendixb}

To incorporate the effect of individual decay of qutrits, we extend the master equation of the main text as follows:
\begin{align}
\label{meq_atomic_decay}
\dot{\hat{\rho}}= & -i [\hat{H},\hat{\rho}] +\kappa\mathcal{L}[\hat c]\hat{\rho} \nonumber \\
& +\sum_{\mu=0,2} \left(\Gamma_{\mu}\mathcal{L}[\hat L_\mu]\hat{\rho}+ \sum_{j=1}^n \Gamma_{1\mu}\mathcal{L}[\hat{A}^j_{1\mu}]\hat{\rho} \right),
\end{align}
where $\hat L_\mu:=\sum_{k=1}^n|\mu\rangle_k\langle1|$ describes the collective decay of qutrits from state $|1\rangle$ to state $|\mu\rangle$ and $\hat{A}_{1\mu}^j= |\mu\rangle_j\langle 1|$ describes the individual decay of qutrits. Due to this additional dissipation channel, the dynamics become more complex. It extends beyond the totally symmetric subspace, thus the semiclassical description is not applicable anymore. Despite these technical difficulties, the qualitative behavior is remarkably similar to the evolution with only bosonic and/or collective qutrit decay. 

\begin{figure}
\centering
\includegraphics[height=45mm]{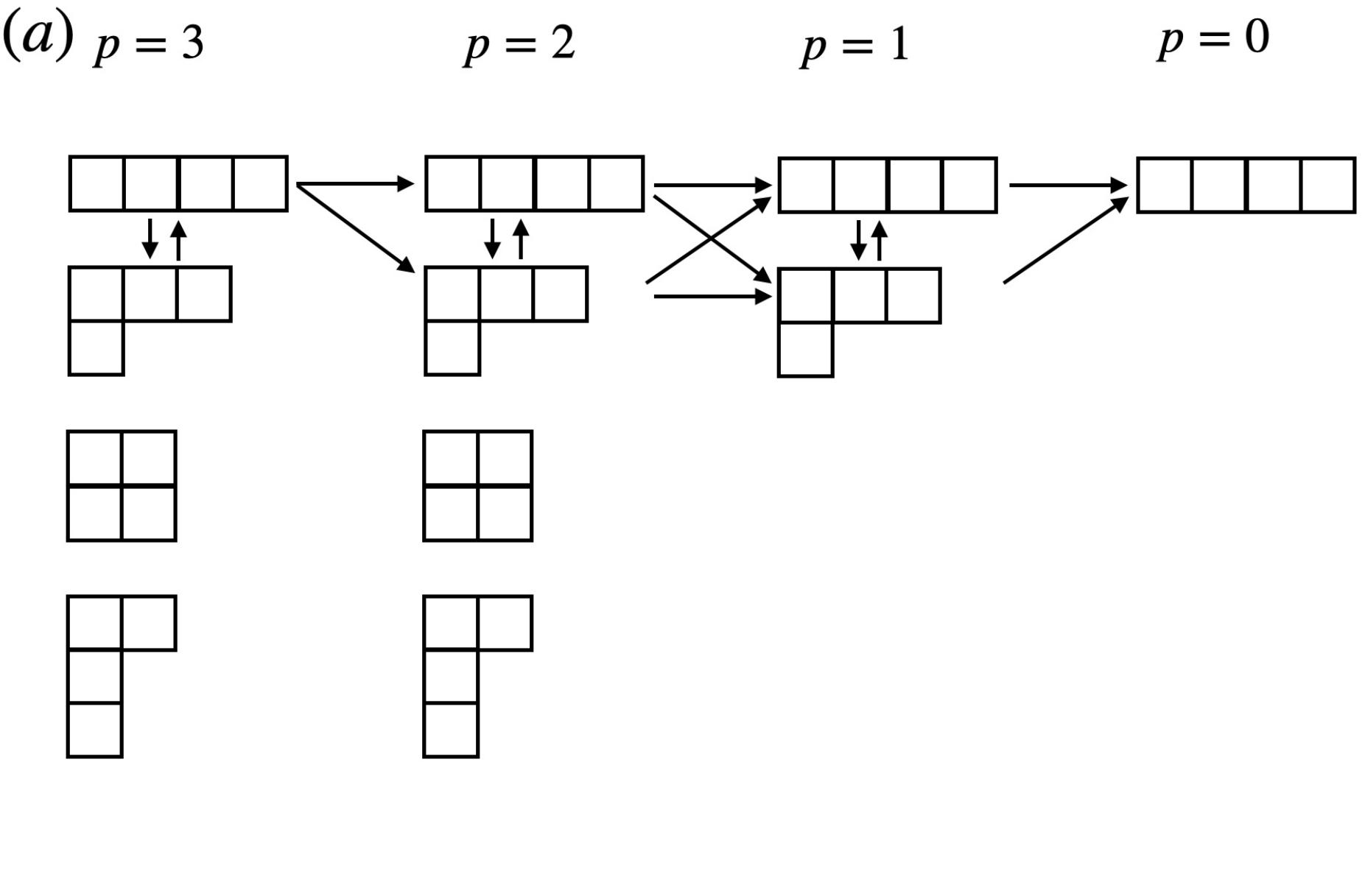} 
\hspace{0.2cm}
\includegraphics[height=45mm]{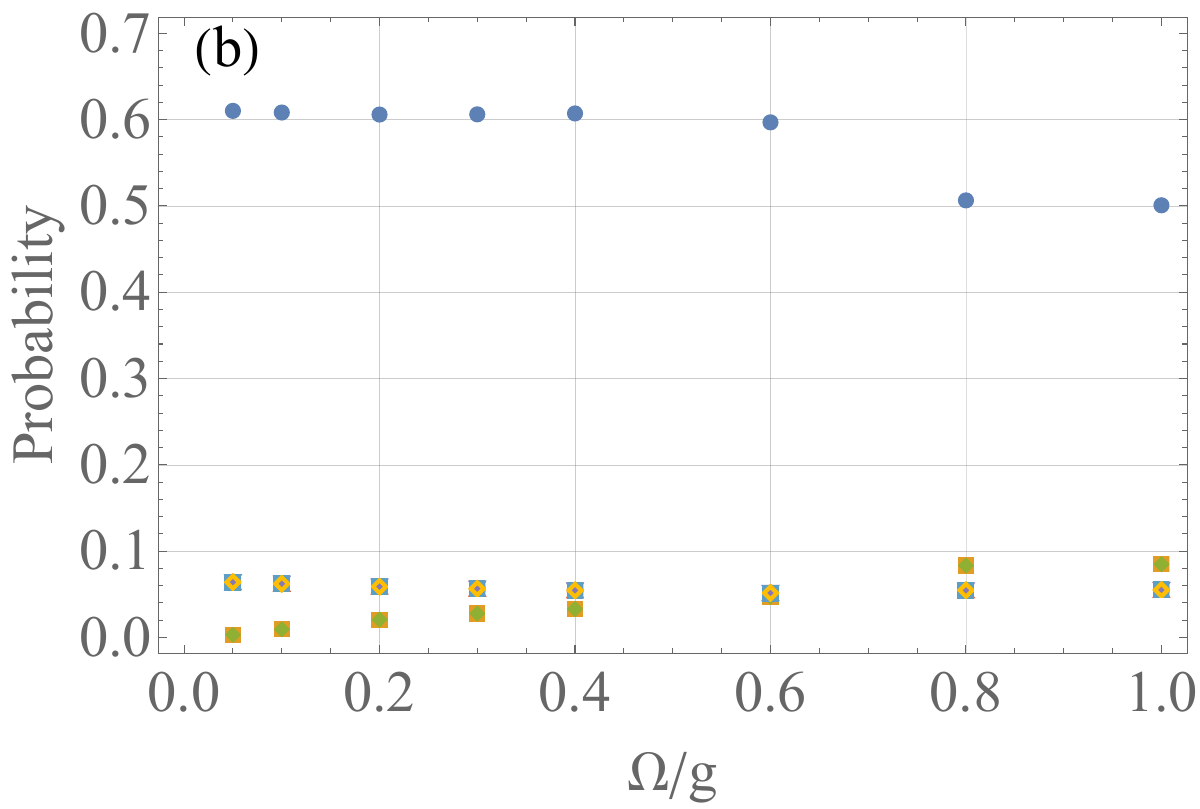}
\caption{Effects of individual decay of qutrits on the initial state $|E_{1;11}\rangle$, for $n=4$. Panel (a) represents the allowed transitions between different symmetry sectors. In panel (b) we show the approximate probability of reaching the various eigenstates with $p=1$, computed as in Eq.~(\ref{eq:Pm}). The eigenstate with the largest $P_m$ (blue dots) is $|E_{1;00}\rangle$.}
\label{Supplemental::fig:3}
\end{figure}

We first only include the individual decay of qutrits ($\kappa=\Gamma_0 = \Gamma_2 =0$). Starting from $|E_{1;11}\rangle$, the allowed transitions between different symmetry sectors (Young diagrams) are represented in Fig.~\ref{Supplemental::fig:3}(a). The individual decay $|1\rangle\to|0\rangle$ gives the `horizontal' transitions, which reduce the number of excitations $p$. On the other hand, the $|1\rangle\to|2\rangle$ decay induces `vertical' transitions, which do not change $p$. If we restrict ourselves to the former process (setting $\Gamma_{12}=0$), after two decays the systems is one of the $|E_1^m\rangle$ eigenstates of the $p=1$ subspace. We can estimate the probabilities of these states as follows:
\begin{align}\label{eq:Pm}
 P_m \simeq  
\sum_k \frac{\Gamma(E_2^k \to E_1^m)}{\sum_{m'} \Gamma(E_2^k \to E_1^{m'})}\frac{\Gamma(E_{1;11} \to E_2^{k})}{\sum_{k'} \Gamma(E_{1;11} \to E_2^{k'})}, 
\end{align}
where $\Gamma(\alpha \to \beta) = \sum_i \Gamma_{10} |\langle \beta|\hat A_{10}^i|\alpha \rangle|^2$ and the sum over $k$ runs over all the intermediate states with $p=2$. In Fig.~\ref{Supplemental::fig:3}(b) we show that the largest $P_m$ is for the state $|E_{1;00}\rangle$, indicating that the decay to the master dark state is the dominant process. Thus, the same type of two-stage dynamics discussed already should take place. 

\begin{figure*}
\centering
\includegraphics[width=4.35cm]{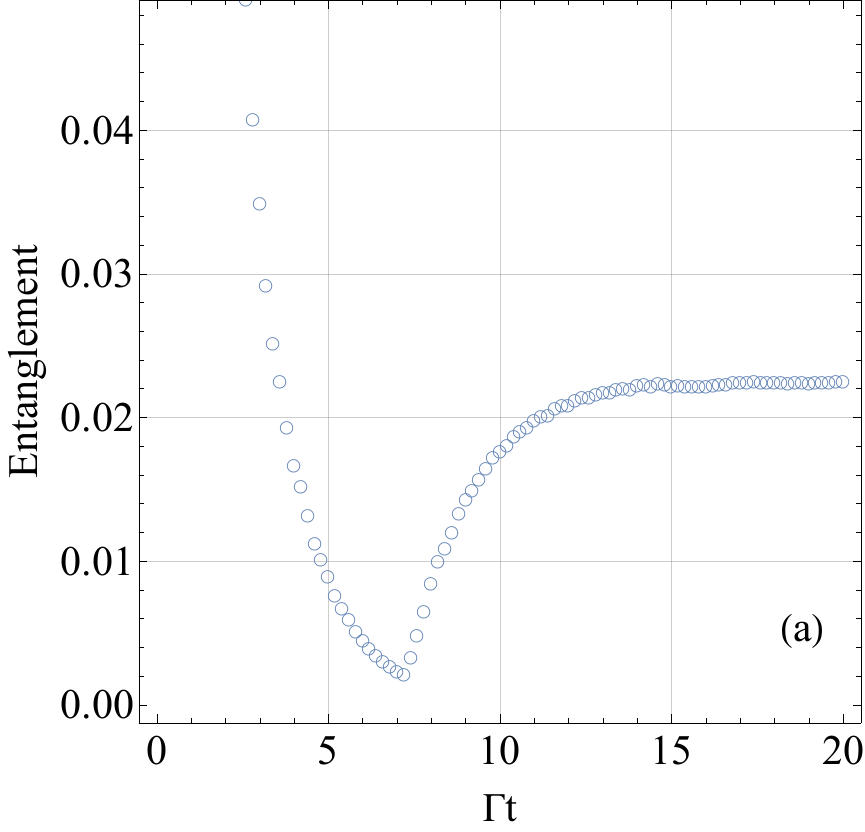}
\hspace{0.2cm}
\includegraphics[width=4.46cm]{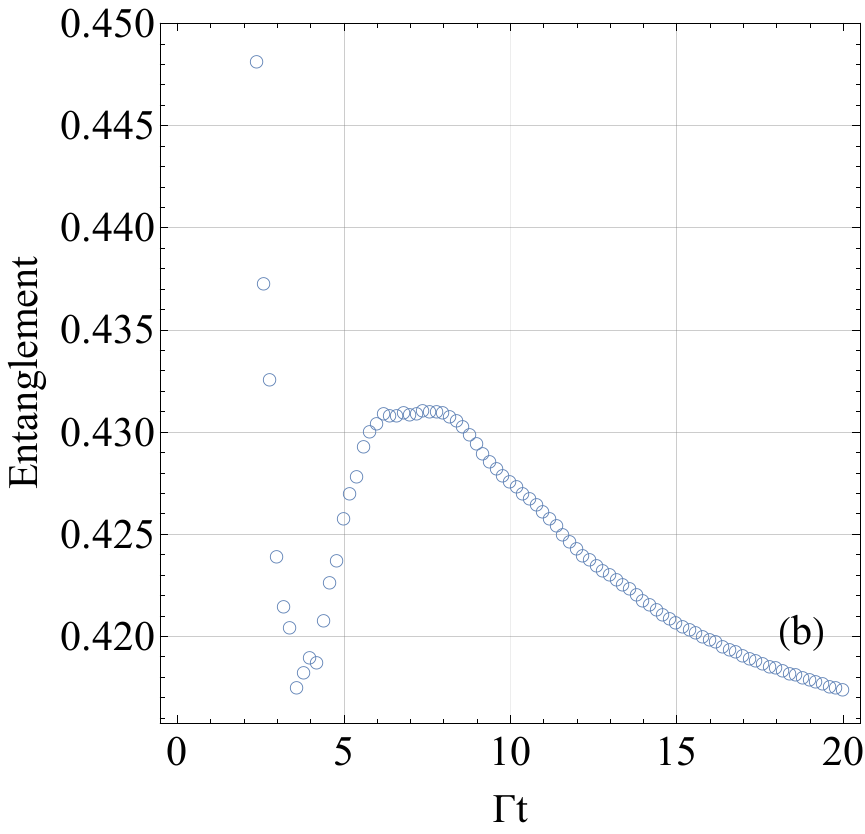} 
\hspace{0.2cm}
\includegraphics[width=4.46cm]{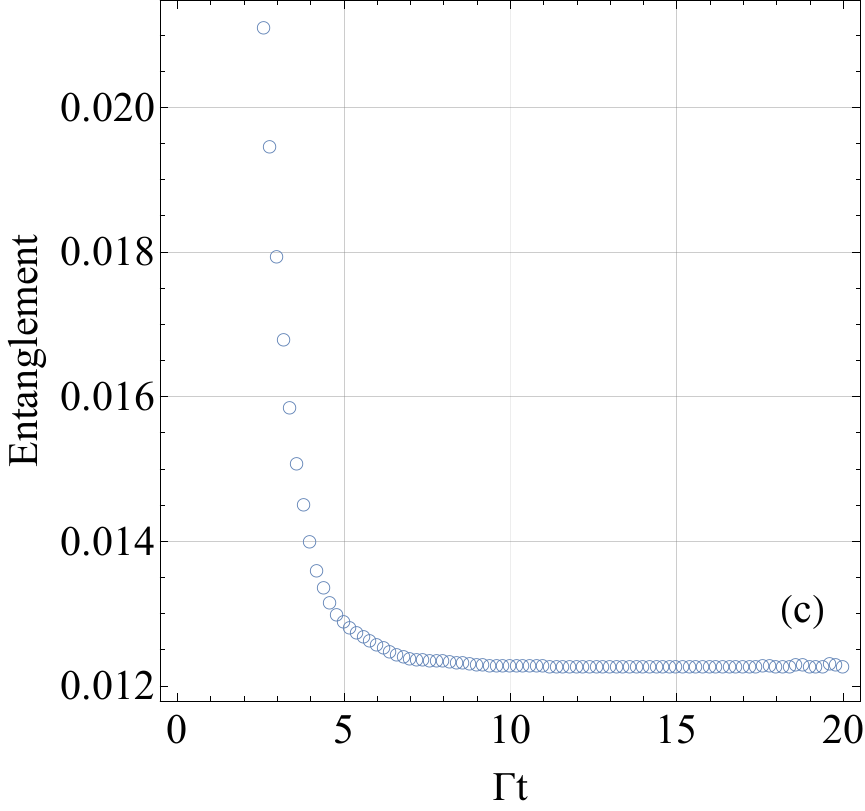}
\caption{Logarithmic negativity simulations with individual qutrit decay. In panel (a) we use $\Gamma_{10}/g=0.05$ and $\Gamma_{12}/2=2.5\times 10^{-3}$. In panel (b) $\Gamma_{10}/g=2.5\times 10^{-3}$ and $\Gamma_{10}/g=0.05$. In panel (c) $\Gamma_{10}=\Gamma_{12}=\Gamma=0.05g$. Other parameters are $n=4$, $p=3$, $\kappa=0$, and $\Omega/g=0.2$.
}
\label{Supplemental::fig:5}
\end{figure*}

In agreement with this argument, we see a clear revival in Fig.~\ref{Supplemental::fig:5}(a), where the decay $|1\rangle \to |0\rangle$ is dominant.  If the $|1\rangle \to |2\rangle$ decay is dominant, the dynamics is also  restricted, as it remains approximately confined to the $p=3$ subspace. As seen Fig.~\ref{Supplemental::fig:5}(b), a strong revival exists in this case. Instead, when the effects of $\hat{A}_{10}^j$ and $\hat{A}_{12}^j$ are comparable, all the processes shown in Fig.~\ref{Supplemental::fig:3}(a) occur on a similar timescale and it is difficult to achieve the self-purification process. As a consequence, the revival is not observed in panel (c) of Fig. \ref{Supplemental::fig:5}.

Lastly, we investigate the entanglement revival when both bosonic decay and individual decay of the qutrits are present. As shown in Fig.~\ref{FigS2}(a), where all the decay rates are comparable, the entanglement revival can persists in this scenario as well.

\section{Detuning}
\label{Appendixc}

\begin{figure*}
\centering
\includegraphics[width=70mm]{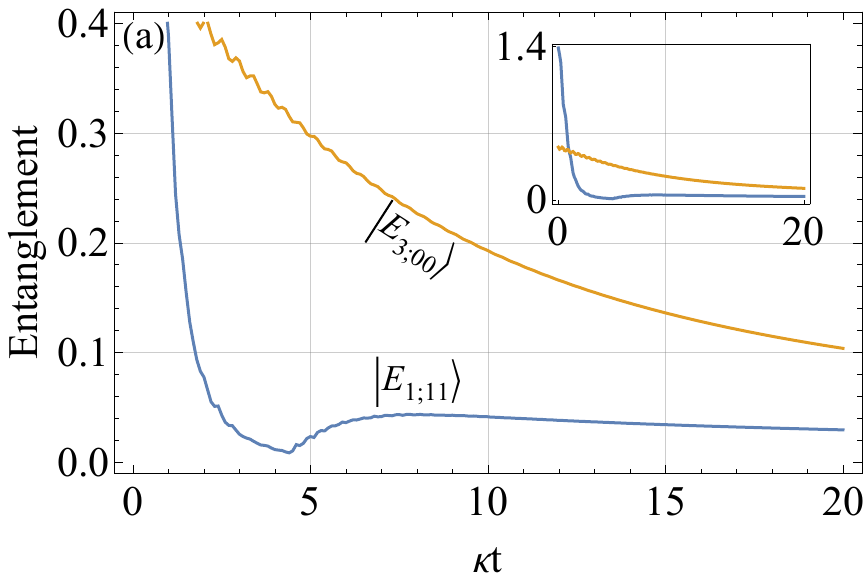} \hspace{0.3cm}
\includegraphics[width=70mm]{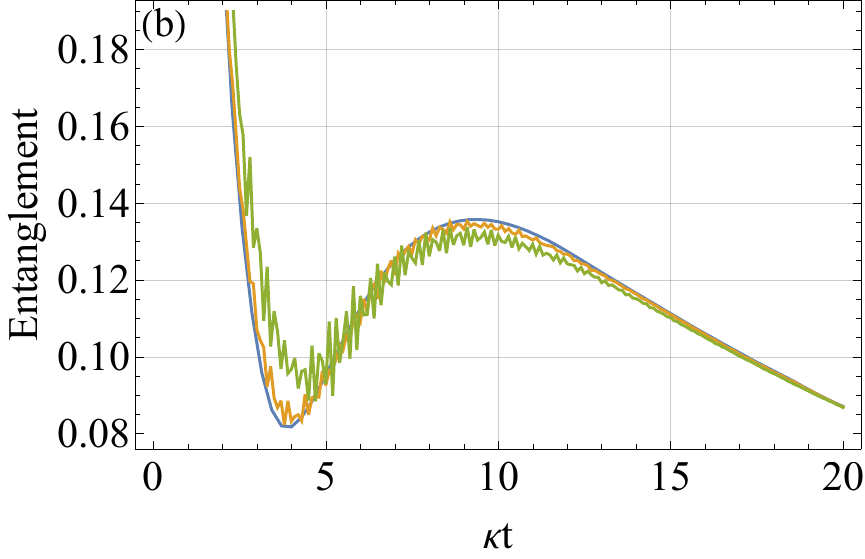}    
\caption{Panel (a): Time dependence of the entanglement including both boson decay and individual qutrit decay. As initial state, we consider either the master dark state $|E_{3;00}\rangle$ (orange curve) or the other zero-energy eigenstate $|E_{1;11}\rangle$ (blue curve). The inset shows the same plot in a larger range. In this panel we use: $\Gamma_{10}/g=2 \Gamma_{12}/g=0.05$, $\Gamma_0=\Gamma_2=0$, and $\Omega/g=0.15$. Panel (b): Entanglement revivals at finite detuning. The three curves refer to $\Delta/(g\sqrt n)=0$ (blue), $0.2$ (orange), and $0.4$ (green). Furthermore we used $\Omega=0.1 g\sqrt{n}$ and neglected qutrit decays. In both panels $n=4$, $p=3$, and $\kappa/g=0.1$.}
\label{FigS2}
\end{figure*}

Until now, we have treated the resonant case. Considering a finite detuning $\Delta$,  the effective Hamiltonian in the bosonic representation reads:
\begin{equation}
    \hatH_{\Delta}=\Delta \hata_1^\dagger\hata+g\sqrt{n}(\hata_1^\dagger\hata_0+\hata_0^\dagger\hata_1)+\Omega(\hata_1^\dagger\hata_2+\hata_2\hata_1^\dagger).
\end{equation}
Simulations for different detunings are presented in Fig.~\ref{FigS2}(b). If the detuning is small compared to the effective coupling strength $g\sqrt n$, the entanglement revival behavior survives because the master dark state $|Z_p^0\rangle$ still satisfies $\hatH_{\Delta}|Z_p^0\rangle=0$. If the detuning is too large, the oscillations caused by detuning and the possibility of transitions from a master dark state $|Z_p^0\rangle$ to other excited states can destroy the entanglement revival behavior.

\section*{References}
\bibliographystyle{iopart-num-long}
\bibliography{References}

\end{document}